\documentclass[prl,aps,twocolumn,floats,showpacs,superscriptaddress]{revtex4}
\usepackage{graphicx}

\newcommand{\be}{\begin{equation}}
\newcommand{\ee}{\end{equation}}
\newcommand{\bea}{\begin{eqnarray}}
\newcommand{\eea}{\end{eqnarray}}

\newcommand{\lp}{\left(}
\newcommand{\rp}{\right)}

\newcommand{\etal}{{\it et al.}\,}

\begin{document}
\title{Interaction effects in the transport of two-dimensional holes in GaAs}
\author{Jian Huang}
\affiliation{%
Department of Electrical Engineering,
Princeton University, Princeton, New Jersey 08544, USA\\}%
\author{D. S. Novikov}
\affiliation{%
Department of Electrical Engineering,
Princeton University, Princeton, New Jersey 08544, USA\\}%
\affiliation{%
W. I. Fine Theoretical Physics Institute, University of Minnesota,
Minneapolis, Minnesota 55455, USA\\}%
\author{D. C. Tsui}
\affiliation{%
Department of Electrical Engineering,
Princeton University, Princeton, New Jersey 08544, USA\\}%
\author{L. N. Pfeiffer}%
\author{K. W. West}%
\affiliation{
Bell Laboratories, Lucent Technologies, Murray Hill, New Jersey 07974, USA}
\date{\today}

\begin{abstract}

The power-law increase of the conductivity with temperature in the
nominally insulating regime, recently reported for the dilute
two-dimensional holes [cond-mat/0603053], is found to systematically
vary with the carrier density. Based on the results from four
different GaAs heterojunction-insulated-gate field-effect-transistor
samples, it is shown that the power law exponent depends on a single
dimensionless parameter, the ratio between the mean carrier
separation
and the distance to the metallic gate that screens the Coulomb
interaction. This dependence suggests that the carriers form a
correlated state in which the interaction effects play a significant
role in the transport properties.

\end{abstract}

\maketitle

Charge transport in two-dimensional (2D) electron systems \cite{AFS}
provides a unique means of studying the interplay between disorder
and electron-electron interactions.
This problem is fundamentally important and remains a subject of
intense investigation \cite{AKS-review}. While noninteracting 2D
electrons are generally believed to form the Anderson insulator
\cite{Anderson'58,stl}, the situation appears to be much more
complex in the presence of interactions \cite{AA,Fin,GD,ZNA}. Recent
theoretical studies \cite{BAA,Pun} emphasize the importance of
collective phenomena around the so-called metal-to-insulator
transition (MIT).

In practice, since the ratio of the Coulomb interaction to the
kinetic energy increases with lowering the density, samples with
most dilute carriers are best suited for probing the collective
phenomena. However, reducing the carrier density $n$ runs into a
risk of increasing the carrier separation $\propto n^{-1/2}$ beyond
the single-particle localization length $\xi$, in which case the
interaction effects become overshadowed by the single-particle
localization. Thus, a sufficiently clean 2D environment is another
requirement that has to be met to uncover the underlying interaction
effects.

The experimental progress on studying the transport of 2D systems
has been greatly influenced by the sample quality. Very early
experimental results in Si-devices demonstrated the
activated transport consistent with the Anderson
insulator \cite{AKS-review}: the Arrhenius conductivity $\sigma \sim e^{-T_A/T}$
at high temperatures, and the softer temperature dependence,
$\sigma\sim e^{-(T^*/T)^\nu}$ at lower temperatures, with $\nu=1/3$ corresponding
to the variable-range hopping (VRH) scenario \cite{Mott-VRH}, and $\nu=1/2$
to the effect of the Coulomb gap \cite{ES}.
However, in the mid-1990's, experiments performed in much cleaner 2D
electrons in Si-MOSFETs showed both metal-like $(d\sigma/dT<0)$ and
insulator-like ($d\sigma/dT>0$) conductivity behavior, depending on
whether the density is above or below a certain critical value $n_c$
\cite{mit}. Although the existence of the metallic regime at $T\to
0$ is still debated, the transport on the insulating side generally
remains activated, in accord with the Anderson localization picture.

The experimental findings on the insulating side of the MIT has
radically changed with the adoption of the undoped GaAs/AlGaAs
heterojunction-insulated-gate field-effect transistors (HIGFETs).
Recent experiments \cite{noh,jian-1} demonstrated that the
conductivity can become non-activated, while preserving the
``insulating'' sign $d\sigma/dT>0$.
In 2003, a close-to-linear dependence $\sigma\propto T$ was first
observed in 2D holes in a $p$-GaAs HIGFET device for densities down
to $p=1.5\times10^{9}$\,cm$^{-2}$ \cite{noh}. Subsequent experiments
in the devices of the same kind not only confirmed this observation
for similar carrier densities, but also revealed a more general
power-law-like temperature dependence, $\sigma\propto T^\alpha$,
with a varying exponent $1\lesssim \alpha\lesssim 2$, at
sufficiently low temperatures \cite{jian-1}. Remarkably, such a
behavior persists even for a record-low density of
$p=7\times10^{8}$\,cm$^{-2}$, in which case the Coulomb
energy is about 100 times greater than the nominal Fermi energy, and
the Fermi wavelength $\lambda_F=\sqrt{2\pi/p} \simeq 0.95\,\mu$m
approaches a macroscopic scale.


In this Letter we present a comprehensive study of this surprising
power law $T$-dependence based on data collected from four different
$p$-type HIGFET samples that only differ by the structural barrier
thickness (the distance $d$ from the 2D layer to the metal gate,
Table \ref{tab:bt}). The procedures of sample preparation and the
measurement details are described in Ref.~\cite{jian-1} for the
first three samples, while the data from the fourth sample is drawn
from Ref.~\cite{noh} for comparison. The fitting of the conductivity
for the temperatures $35 < T\lesssim 200\,$mK to \be
\label{power-law} {\sigma / \sigma_Q}  = G_0 + \lp {T/ T_0}
\rp^\alpha \,, \quad \sigma_Q = e^2/(2\pi \hbar) \ee yields a
sample-dependent exponent $\alpha(p)$ that grows with decreasing
hole density $p$, while the $T$-independent term remains negligible,
$G_0\sigma_Q \ll \sigma(T)$. Moreover, we find that $\alpha$
systemically depends on the single dimensionless parameter $\kappa =
a/d$, where $a=(\pi p)^{-1/2}$ is the Wigner-Seitz radius (the mean
carrier distance is about $2a$). Such a single-parameter dependence,
tied to the screening length $\simeq 2d$ for the Coulomb interaction
rather than to the sample-specific localization length, strongly
indicates an important role played by the electron interactions.

\begin{table}[b]
\caption{\label{tab:bt} The barrier thickness $d$ (distance to the
metallic gate) for four different samples}
\begin{ruledtabular}
\begin{tabular}{lccccc}
Samples\ & \#2 \ & \#3 \ & \#4 \ & Noh \etal \\
Barrier $d$\ & 600\,nm\ & 600\,nm\ & 250\,nm\ & 500\,nm\\
\end{tabular}
\end{ruledtabular}
\end{table}


\begin{figure}[t]
\includegraphics[totalheight=4.5in]{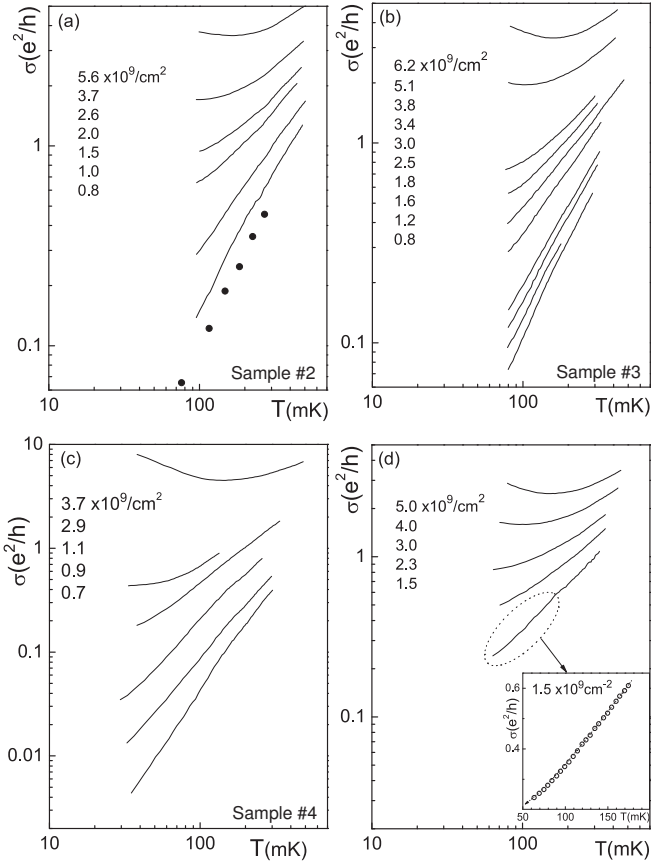}
\caption{\label{fig:ctlg}
Conductivity temperature dependence in the log-log scale
for a set of hole densities, for the samples
(a) \#2, (b) \#3, (c) \#4, and (d) the sample from Ref.~\cite{noh}.
The scattered points in (b) are the dc results for $p=8\times10^{8}$\,cm$^{-2}$.
}
\end{figure}


The aim of this work is to analyze the power-law-like temperature
dependence  $\sigma(T)$ and to examine the relevance of the
interaction effects. In Fig.~\ref{fig:ctlg}, we show the log-log
plots of $\sigma(T)$ obtained from four different samples: (a) \#2
and (b) \#3 are samples from the same wafer, cooled down to the
lowest temperature of around 80\,mK; (c) sample \#4 is cooled down
to 35\,mK; and (d) is the data from Ref.~\cite{noh} with the lowest
$T$ of 65\,mK. Each panel in Fig.~\ref{fig:ctlg} contains curves for
a number of hole densities. For the density above critical, $p_c
\simeq 4\times10^{9}$\,cm$^{-2}$, the transport is metal-like
\cite{AKS-review}. Below we focus on the opposite, low-density
``insulating'' regime.

There are three features common to all four samples. (i)
Although the sign
$d\sigma/dT>0$, the conductivity never becomes activated
\cite{jian-1} because an activated $T$-dependence in a log-log plot
would have shown a strong downward bending for $T$ below the
activation temperature.
(ii) For low densities $p<2\times10^{9}$\,cm$^{-2}$, the slopes
$d\log\sigma/d\log T$ are roughly temperature-independent at
temperatures $\lesssim 200\,$mK and increase with decreasing
density. (iii) Finally, although the non-activated conductivity is a
signature of the extended states, the conductivity values are 1-2
orders below $e^2/(2\pi\hbar)$ (the 2D analog of the minimal
metallic conductivity \cite{Mott-min-met}).


\begin{figure}[b]
\includegraphics[height=5.7in]{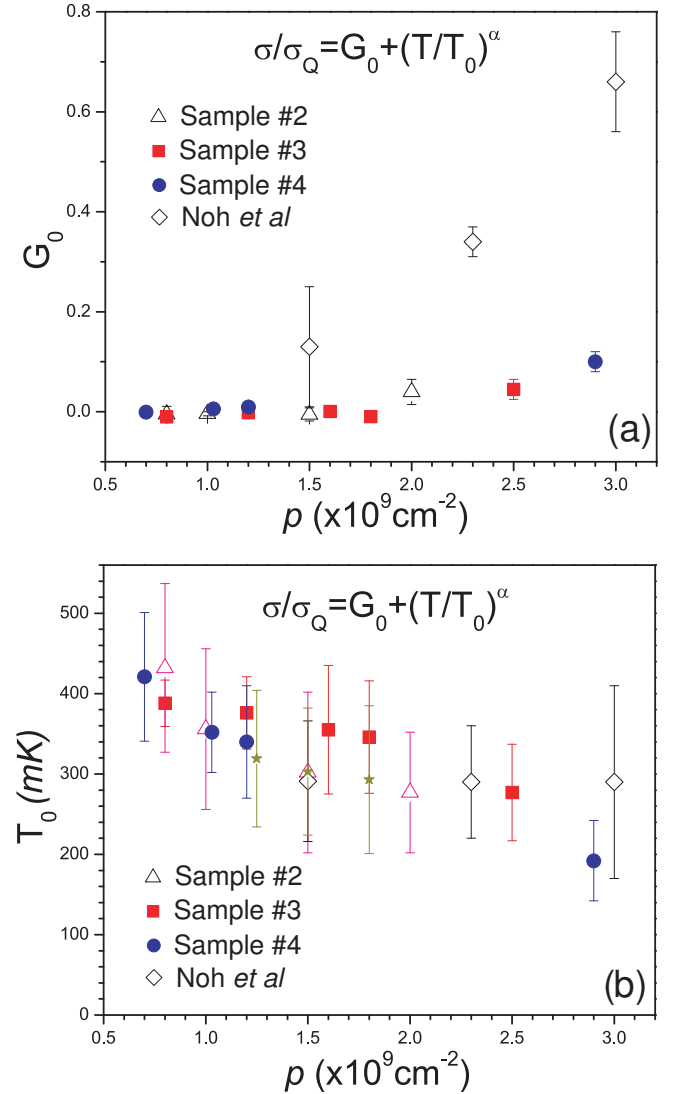}
\caption{\label{fig:ab} Density dependence of the fitting parameters
$G_0$ and $T_0$ from Eq.~(\ref{power-law})
}
\end{figure}

\begin{figure}[b]
\includegraphics[width=3.3in, trim=0.01in 0.05in 0.02in 0.03in]{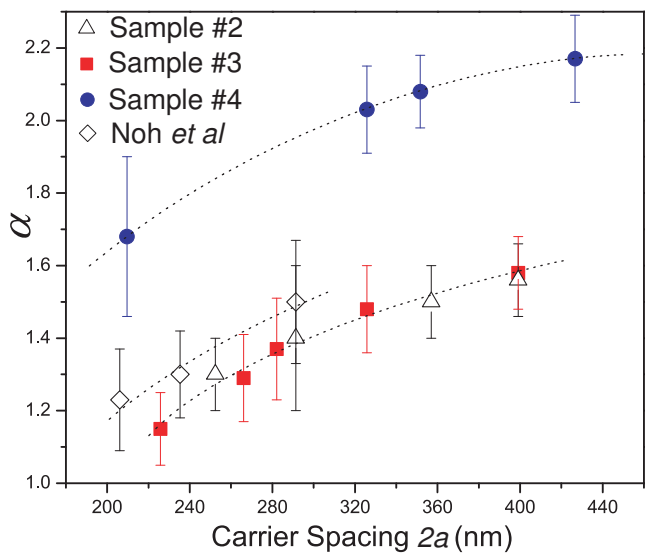}
\caption{\label{fig:expo-2a} Density dependence of the exponent
$\alpha$ for all the samples.
The dotted lines are guides for the eye.}
\end{figure}

In what follows we investigate the density dependence of the slope
$\alpha = d\log\sigma/d\log T$ by fitting $\sigma(T)$ to the simple
formula (\ref{power-law}) for the low density curves, $p \leq
3\times10^{9}$\,cm$^{-2}$. The fitting parameters $G_0$ and $T_0$
are plotted in Fig.~\ref{fig:ab}. The $T$-independent term $G_0$
remains around zero up to $p \simeq 2\times10^{9}$\,cm$^{-2}$ for
the samples \#2, \#3, and \#4, indicating a fairly good power law
dependence: $\sigma \propto T^\alpha$. On the other hand, for the
sample from Ref.~\cite{noh}, the constant $G_0$ is more significant.
We stress that, although the high-temperature behavior in the sample
from Noh \etal looks quite linear, $\sigma \simeq A + BT$, as
reported in Ref.~\cite{noh}, at $T\lesssim 200\,$mK the conductivity
curve exhibits a noticeable bending at $T<200\,$mK as demonstrated
in the linear scale plot in the inset of Fig.\ref{fig:ctlg} (d), and
is best fitted to Eq.~(\ref{power-law}) with $\alpha\neq 1$ [see
Fig.~\ref{fig:expo-2a}].
The parameter $T_0$ for all four samples has a trend of a slow
decrease with increasing density from about 400\,mK to about 300\,mK
for densities up to $p \simeq 2\times10^{9}$\,cm$^{-2}$.

Fig.~\ref{fig:expo-2a} shows how the exponent $\alpha$ depends on
the parameter $2a$, which is approximately the mean carrier spacing.
The results from samples \#2 and \#3 fall approximately onto a
single curve in which $\alpha$ varies from 1.1 to about 1.5. The
results from both samples \#4 and Ref.~\cite{noh} qualitatively
follow the same trend albeit the values of $\alpha$ are greater by
about 0.6 and 0.1 respectively.
%
This shift in $\alpha(p)$ relative to that for samples \#2 and \#3
motivates us to look into the structural differences between the
samples, which is primarily in the barrier thickness $d$.
As shown in Table \ref{tab:bt}, samples \#2 and \#3 have the same
$d$, while the values of $d$ for the other two samples are notably
different. Furthermore, the values of $\alpha$ increase with the
decrease in $d$ for a given density. This trend points at the role
of the screening by the metallic gate.
%

\begin{figure}[t]
\includegraphics[width=3.3in, trim=0.07in 0.05in 0.02in 0in]{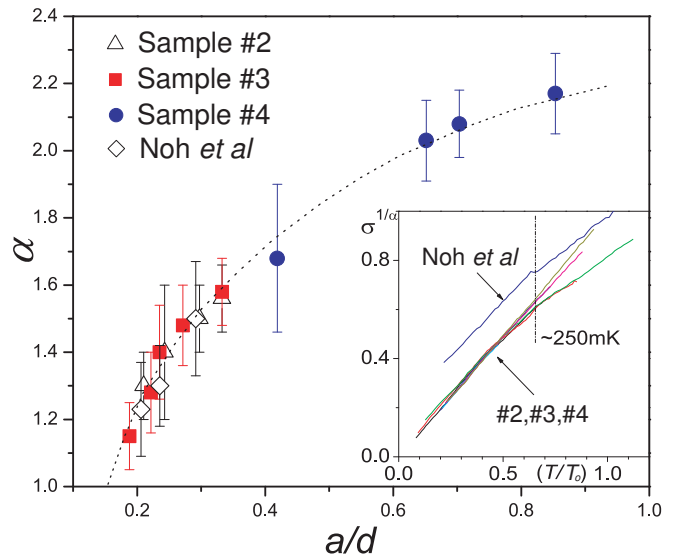}
\caption{\label{fig:all}
The exponent $\alpha$ as a function of the ratio $\kappa = a/d$.
Inset: $\sigma^{1/\alpha}$ as a function of $T/T_0(p)$.
}
\end{figure}


In HIGFETs, the metallic gate at the distance $d$ from the 2D hole layer
screens the $1/r$ interaction down to $1/r^3$ when $r\gtrsim 2d$.
For the lowest densities, the interaction becomes effectively
short-ranged, with the relevant parameter being the ratio $\kappa =
a/d$ between the carrier spacing $\approx 2a$ and the screening
radius $2d$. For our measurements, this ratio can be continuously
varied in the range $0.1 < \kappa < 0.85$; the sample of Noh \etal
also falls into this range. The effect of the gate screening becomes
apparent in Fig.~\ref{fig:all}, where all the power law exponents
$\alpha$ are plotted as a function of $\kappa$. Remarkably,
$\alpha(\kappa)$ from all four samples fall onto a single curve
within reasonable error bars. This curve tends to saturate for
$\alpha \gtrsim 2$, while it most strongly varies when $\alpha \approx 1$.
Thus the linear dependence
$\sigma \propto T$ reported in Ref.~\cite{noh} is most probably a crossover
into an entirely different transport regime, rather than a universal signature
at low $T$.
[In the inset we plot $\sigma^{1/\alpha}$ as a function of $T/T_0(p)$
to illustrate the relative insignificance of the constant $G_0$ in
Eq.~(\ref{power-law}).]


The strong sensitivity of the transport to the shape of the
interaction potential is not entirely surprising. Indeed, the
carriers at these densities are delocalized and are very strongly
interacting (the $r_s$ value for $p=1\times 10^9\,$cm$^{-2}$ is 100
if one assumes the hole band mass $m=0.4 m_e$). A plausible way to
think about such a system is by imagining a liquid or a
strongly-interacting plasma. As it has been recently shown, the
kinetic and thermodynamic properties of classical plasmas strongly
depend on the same screening parameter $\kappa$ \cite{plasma},
although the classical arguments alone cannot explain the peculiar
dependence (\ref{power-law}). It is interesting whether the quantum
effects could manifest themselves at lower (so far inaccessible)
temperatures, in which case the system could become collectively
localized \cite{BAA}, or whether the MIT takes place \cite{Pun}. The
dependence on $\kappa=a/d$, where $a$ and $d$ can be independently
controlled, suggests that,
by varying the density, one can continuously modify the state of the system. For
sufficiently low densities ($a\gtrsim d$), a possibility of a reentry into the
Fermi-Liquid (FL) was suggested \cite{spivak2}.
For larger densities, a sequence of mixed phases \cite{spivak} was conjectured.

We finally note that varying $d$ not only changes the interaction
range, but also the correlation length $\xi_{\rm dis}$ of the
disorder potential. The scattering off the surface imperfections at
the gate level is estimated to be negligibly small. In the case when
the disorder is dominated by the charge impurities in the bulk, the
gate screens the disorder potential harmonics for length scales
$\gtrsim d$, so that $\xi_{\rm dis} \sim d$. In such a situation it
becomes more difficult to differentiate the electron-electron
interaction effects from those due to the change in the distribution
of disorder \cite{EPB,Fogler2004}. The dependence of the
conductivity on $a/d$, rather than on $a$ and $d$ separately,
suggests that the electron interactions are probably more important
than the change in disorder. One natural possibility for the
electron interaction to enter is through the screening of the
impurity field (such as via the RPA screening at smaller $r_s$
\cite{GD,DSH}), in which case the resulting effective disorder would
depend on $a/d$ as long as $a<d$. Another, equally plausible
scenario is provided by assuming the hydrodynamic (viscous) flow of
an electron liquid past the impurities whose size $\xi_{\rm dis} >
a$. In this case the resistivity is proportional to the viscosity of
the 2D liquid \cite{HruskaSpivak,spivak,SK_annphys} (that depends on
$\kappa$), and to the number of impurities, while the dependence on
their shape and size (that may become affected by varying $d$)
enters only under the logarithm, according to the well-known Stokes
paradox of a 2D flow \cite{lamb}.

In summary, by performing transport measurements on high quality 2D
holes with densities down to $7\times 10^8\,$cm$^{-2}$, we
established the dependence of the power-law exponent $\alpha$
[Eq.~(\ref{power-law})] on the ratio $a/d$ between the Wigner-Seitz
radius $a$ and the distance to the metal gate $d$. We ascribe this
dependence to the screening of the Coulomb interaction by the gate.
We believe that our results provide direct evidence of the role of
the electron-electron interaction in the 2D transport, and suggest
that the transport is sensitive to the shape of the interaction
potential controlled by the screening distance $d$. By varying the
ratio $a/d$, one can realize a strongly-correlated state of the 2D
carriers with tunable properties.

This work has benefited from valuable discussions with I.L. Aleiner,
B.L. Altshuler, R.N. Bhatt, and M.I. Dykman. The work at Princeton
University is supported by US DOE grant DEFG02-98ER45683, NSF grant
DMR-0352533, and NSF MRSEC grant DMR-0213706.
The work at the FTPI is supported by NSF grants DMR 02-37296 and DMR 04-39026.

\end{document}